\documentclass[aps,reprint,prl,superscriptaddress,showpacs,floats,floatfix]{revtex4-2}
\usepackage{morefloats}
\usepackage{mathptmx}
\usepackage{indentfirst}
\usepackage{amsmath,graphicx,dcolumn}
\usepackage[colorlinks,citecolor=green]{hyperref}
\usepackage[usenames]{color}
\usepackage{datetime}
\usepackage{textcomp}
\usepackage{ulem}
\usepackage{float}

\newcommand{\blue}{\textcolor{blue}}

\newcommand*{\LNO}{La$_3$Ni$_2$O$_7$}
\newcommand*{\mjcm}{$\mu$J/cm$^2$}
\begin{document}

\title{Ultrafast Optical Evidence of Coexisting Density Waves in Bilayer Nickelate {\LNO}}

\author{Qi-Yi Wu}
\thanks{These authors contributed equally to this work}
\affiliation{School of Physics, Central South University, Changsha 410012, Hunan, China}
\affiliation{School of Materials Science and Engineering, Central South University, Changsha 410083, Hunan, China}

\author{De-Yuan Hu}
\thanks{These authors contributed equally to this work}
\affiliation{Center for Neutron Science and Technology, Guangdong Provincial Key Laboratory of Magnetoelectric Physics and Devices, School of Physics, Sun Yat-Sen University, Guangzhou 510275, Guangdong, China}

\author{Chen Zhang}
\affiliation{School of Physics, Central South University, Changsha 410012, Hunan, China}

\author{Mengwu Huo}
\affiliation{Center for Neutron Science and Technology, Guangdong Provincial Key Laboratory of Magnetoelectric Physics and Devices, School of Physics, Sun Yat-Sen University, Guangzhou 510275, Guangdong, China}

\author{Hao Liu}
\affiliation{School of Physics, Central South University, Changsha 410012, Hunan, China}

\author{Bo Chen}
\affiliation{School of Physics, Central South University, Changsha 410012, Hunan, China}

\author{Ying Zhou}
\affiliation{School of Physics, Central South University, Changsha 410012, Hunan, China}

\author{Zhong-Tuo Fu}
\affiliation{School of Physics, Central South University, Changsha 410012, Hunan, China}

\author{Chun-Hui Lv}
\affiliation{School of Physics, Central South University, Changsha 410012, Hunan, China}

\author{Zi-Jie Xu}
\affiliation{School of Physics, Central South University, Changsha 410012, Hunan, China}

\author{Hai-Long Deng}
\affiliation{School of Physics, Central South University, Changsha 410012, Hunan, China}

\author{H. Y. Liu}
\affiliation{Beijing Academy of Quantum Information Sciences, Beijing 100085, China}

\author{Jun Liu}
\affiliation{School of Materials Science and Engineering, Central South University, Changsha 410083, Hunan, China}

\author{Yu-Xia Duan}
\affiliation{School of Physics, Central South University, Changsha 410012, Hunan, China}

\author{Meng Wang}
\email{Corresponding author: wangmeng5@mail.sysu.edu.cn}
\affiliation{Center for Neutron Science and Technology, Guangdong Provincial Key Laboratory of Magnetoelectric Physics and Devices, School of Physics, Sun Yat-Sen University, Guangzhou 510275, Guangdong, China}

\author{Jian-Qiao Meng}
\email{Corresponding author: jqmeng@csu.edu.cn}
\affiliation{School of Physics, Central South University, Changsha 410012, Hunan, China}

\date{\today}

\begin{abstract}
Utilizing ultrafast optical pump-probe spectroscopy, we investigate the coexistence and competition of electronic orders in the bilayer nickelate {\LNO}. Our results reveal two coexisting density waves that can be selectively manipulated with light. We directly identify a spin-density wave (SDW) with electronic nematicity emerging below $T_{\rm SDW}$ $\approx$ 140 K by measuring its spin dynamics, and discover a distinct, nonmagnetic charge order appearing below $T_{\rm DW}$ $\approx$ 115 K. The central finding is the demonstration of differential optical control: the charge order is fragile, completely suppressed by a pump fluence of  approximately 40 {\mjcm}, while the SDW is remarkably robust, persisting to 200 {\mjcm}. This work establishes a clear hierarchy in the stability of competing electronic orders and provides a powerful method for disentangling their interplay in quantum materials.

\end{abstract}
\maketitle

The recent discovery of superconductivity in {\LNO} at 14 GPa, exhibiting a critical temperature ($T_c$) of approximately 80 K \cite{HSun2023}, has profoundly stimulated research into high-temperature superconductors \cite{YNZhang2024, EKKo2025, GZhou2025, LCRhodes2024, ZLuo2023, VChristiansson2023, FLechermann2023, QQin2023, QGYang2023, KJiang2024, JYYou2025, XZQu2024, KYJiang2025, JZhan2025}. In the context of high-temperature superconductivity, antiferromagnetic spin fluctuations are widely considered to be a crucial mediator of superconducting pairing. Within substantial regions of the phase diagram, intricate interactions among spin, charge, and lattice degrees of freedom often lead to exotic ordering phases, such as charge density wave (CDW) in cuprates \cite{SMHayden2024, EHda2014, SGerber2015} and spin density wave (SDW) in iron-based superconductors \cite{JMAllred2016, PCDai2015, FKretzschmar2016, EFradkin2015}. These orders are of significant interest because of their intimate connection with the mechanisms underpinning high-temperature superconductivity.\setlength{\parskip}{0pt}

Specifically for {\LNO}, theoretical predictions suggest the presence of a SDW order at wave vector ($\pm$$\pi/2$, $\pm$$\pi/2$) and a CDW order at ($\pm$$\pi$, $\pm$$\delta$) with a small $\delta$ \cite{ZLuo2023, VChristiansson2023}. A series of experimental studies have consistently identified a density wave (DW)-like transition occurring between 130-150 K at ambient pressure \cite{YNZhang2024, KChen2024, RKhasanov2025, GWu2001, ZLiu2023, MKakoi2024, DZhao2025, XChen2024, NKGupta2025, XLRen2025}. Evidence from techniques such as muon spin rotation ($\mu$SR) \cite{KChen2024, RKhasanov2025}, nuclear magnetic resonance (NMR) \cite{MKakoi2024, DZhao2025}, resonant inelastic x-ray scattering (RIXS) \cite{XChen2024}, resonant x-ray scattering \cite{NKGupta2025, XLRen2025} points to an SDW origin for this transition. Additionally, optical conductivity measurements have revealed a distinct gap opening below 115 K \cite{ZLiu2024}, which is widely believed to be associated with the formation of a CDW order. High-pressure experiments have further demonstrated that the DW orders in {\LNO} are markedly influenced by increasing pressure \cite{KChen2024, RKhasanov2025, YNZhang2024, GWang2024, YZhou2025, JLi2025}. These collective experimental findings highlight that {\LNO} possesses a rich and intricate landscape of DW orders, thereby establishing it as a promising new platform for unraveling the fundamental relationship between DW order and superconductivity.\setlength{\parskip}{0pt}

Ultrafast optical pump-probe spectroscopy serves as a powerful technique for investigating the interactions among various microscopic degrees of freedom within materials and for characterizing gap dynamics near the Fermi level. This method has been widely adopted in studies of cuprate and iron-based superconductors \cite{Kabanov1999, MCWang2018, EEMChia2010, CWLuo2017}. Following the discovery of nickelate superconductors, ultrafast optical spectroscopy has been swiftly applied to elucidate their properties \cite{YMeng2024, SXu2025, YLi2025, BCheng2024}. This has yielded crucial information pertinent to understanding the mechanisms of high-temperature superconductivity, encompassing insights into the origin and pressure-dependent evolution of density wave orders in {\LNO} \cite{YMeng2024} and La$_4$Ni$_3$O$_{10}$ \cite{SXu2025}, the nature of electron-phonon ($e$-$p$h) coupling in both compounds \cite{YLi2025}, and evidence supporting \textit{d}-wave superconductivity in infinite-layer nickelates \cite{BCheng2024}. Furthermore, intense photoexcitation can transiently reconstruct free-energy landscapes by suppressing order parameters, thereby successfully inducing nonequilibrium phase transitions suppressing order parameters, thereby successfully inducing nonequilibrium phase transition \cite{ATorre2021}. Consequently, ultrafast lasers are rapidly emerging as an important new dynamic tuning tool for manipulating quantum materials \cite{TDong2022}, complementing conventional tuning parameters such as temperature, doping, pressure, and magnetic fields.

\begin{figure*}[t]
\vspace*{-0.2cm}
\begin{center}
\includegraphics[width=2\columnwidth]{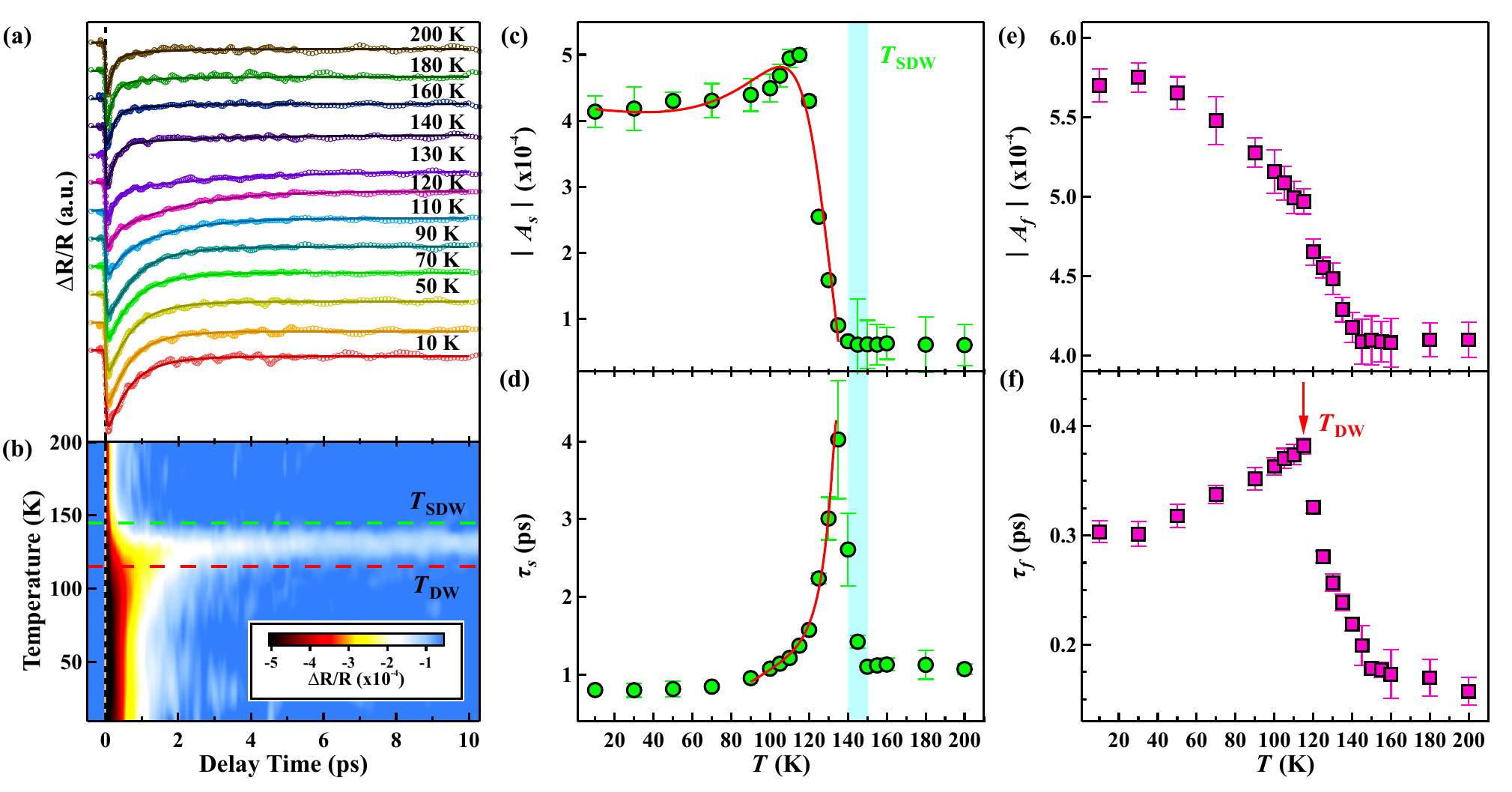}
\end{center}
\vspace*{-0.7cm}
\caption{(a) Transient reflectivity ($\Delta R/R$) as a function of delay time at various temperatures, measured at a pump fluence of $\sim$9.9 {\mjcm}. The experiential data can be well fitted by two-exponential decays. The solid lines are the fitting curves. (b) 2D false-color image of $\Delta R/R$ as a function of temperature and delay time. Red and green dashed lines indicate $T_{DW}$ and $T_{\rm SDW}$, respectively. (c), (d) Temperature dependence of amplitude ($A_s$) and relaxation time ($\tau_s$), respectively. The red solid lines are the RT model fitting curves. (e), (f) Temperature dependence of amplitude ($A_f$) and relaxation time ($\tau_f$), respectively. Error bars are the standard error in the exponential fitting.}
\label{FIG:1}
\end{figure*}

In this Letter, we present a systematic investigation into the nature and evolution of DW-like orders in {\LNO}, employing ultrafast pump-probe spectroscopy. Analysis of quasiparticle dynamics indicates a gap opening below approximately 140 K, attributable to a DW-like order that time-resolved ellipticity measurements definitively identify as an SDW order. Additionally, a distinct suppression of the initial relaxation time of $\Delta R/R$ at approximately 115 K points to the presence of another potential DW-like order. The temperature-fluence phase diagram constructed from our data illustrates that the lower-temperature DW order is significantly suppressed, vanishing above approximately 40 {\mjcm}. In contrast, $T_{\rm SDW}$ is suppressed from approximately 140 K at $\sim$9.9 {\mjcm} to approximately 70 K at $\sim$198.9 {\mjcm}. Concurrently, the SDW gap size is observed to decrease from approximately 69 meV to 15 meV. Collectively, our findings elucidate the fundamental nature of the DW-like order in {\LNO} and provide critical experimental evidence that will inform future investigations into the optical control of density wave phenomena in this material.

The single crystals of {\LNO}, characterized by well-defined (001) cleavage planes, were synthesized using a vertical optical image floating-zone method \cite{HSun2023}. To investigate the ultrafast dynamics of {\LNO} across a temperature range 10-300 K, we employed ultrafast laser pulses with an 800 nm central wavelength, 35 fs pulse width, and a 1 MHz repetition rate. Spin dynamics were investigated through ultrafast ellipticity measurements \cite{MCWang2016, APatz2014}. Except for the ultrafast ellipticity measurements, which used a 400 nm pump laser, all other measurements utilized an 800 nm pump laser. All measurements were conducted on freshly cleaved surfaces under a vacuum of 10$^{-6}$ mbar. Detailed descriptions of all measurement procedures can be found in Sec. \blue{I} of the Supplemental Material \cite{SUPPM} and in previous publications \cite{BLTan2025, QYWu2023, QYWu2025}.

Figure \blue{1(a)} presents the typical transient reflectivity signal $\Delta R/R$ measured at various selected temperatures with a low-pump fluence of approximately 9.9 {\mjcm}. At all temperatures investigated, photoexcitation induces an immediate decrease in reflectivity, followed by a recovery to a constant offset through two distinct relaxation processes, each characterized by its own relaxation time. Figure \blue{1(b)} illustrates a two-dimensional (2D) false-color image of $\Delta R/R$ as a function of both delay time and temperature. A notable observation is the significant change in transient reflectivity near 140 K, particularly evident in the signal beyond 1 ps. As the temperature decreases, $\Delta R/R$ within the time range 0.1- 0.8 ps exhibits discernible anomalies around 115 K. For quantitative study of the quasiparticle relaxation, we fitted the data with a two-exponential decay convoluted with a Gaussian laser pulse $G(t)$ (FWHM $\simeq$ 60 fs) (see Sec. \blue{II} of the Supplemental Material \cite{SUPPM} for details), 
\begin{equation}
\begin{split}
\frac{\Delta R(t)}{R} = [(1-e^{-t/t_{b}})(A_{f}e^{-t/\tau_{f}}+A_{s}e^{-t/\tau_{s}})+C]\otimes G(t) ,
\end{split}
\end{equation}
Here, $A_f$ and $A_s$ represent the amplitudes of the fast and slow components, respectively, while $\tau_f$ and $\tau_s$ denote their corresponding relaxation times, $t_b$ is the build-up time of the non-equilibrium quasiparticle \cite{WZhang2016}, and $C$ is the constant offset. The solid lines in Fig. \blue{1(a)} demonstrate that the fitting curves are in excellent agreement with the experimental data.

The extracted values for the amplitude $A_s$ and relaxation time $\tau_s$ as functions of temperature are summarized in Figs. \blue{1(c)} and \blue{1(d)}, respectively. As the temperature decreases, $A_s$ exhibits a steady increase below $T_{\rm SDW} \sim$140 K (regarded as the onset of the SDW order, which will be discussed later). Concurrently, the relaxation time $\tau_s$ shows a significant increase, appearing to diverge near $T_{\rm SDW}$.  These observed behaviors are commonly reported in CDW compounds \cite{TDong2022, RSLi2022} and Iron-based superconductors with SDW order \cite{TMertelj2010, LStojchevska2012}, suggesting the opening of a DW-like gap.  This phenomenon can be interpreted within the framework of the Rothwarf-Taylor (RT) model in the boson bottleneck regime \cite{ARothwarf1967, JDemsar2006}, as discussed further in Sec. \blue{III} the Supplemental Material \cite{SUPPM}. Figures \blue{1(c)} and \blue{1(d)} illustrate a good fit of the RT model to the experimental data. This fitting yields an energy gap of $\Delta_{\rm SDW}$(0) = 69 $\pm$ 4 meV (see Sec. \blue{III} of the Supplemental Material \cite{SUPPM} for details). The gap size determined in this study aligns well with values previously reported through NMR studies \cite{MKakoi2024, DZhao2025}, and other pump-probe experiments \cite{YMeng2024}. It is important to note that the transition temperature $T_{\rm SDW}$ presented here is slightly lower than the 150 K reported in some prior experiments \cite{KChen2024, RKhasanov2025, ZLiu2023}. This discrepancy might be attributable to the inhomogeneous nature of {\LNO} \cite{XSNi2025}.

\begin{figure}[t]
	\vspace*{-0.2cm}
	\begin{center}
		\includegraphics[width=1\columnwidth]{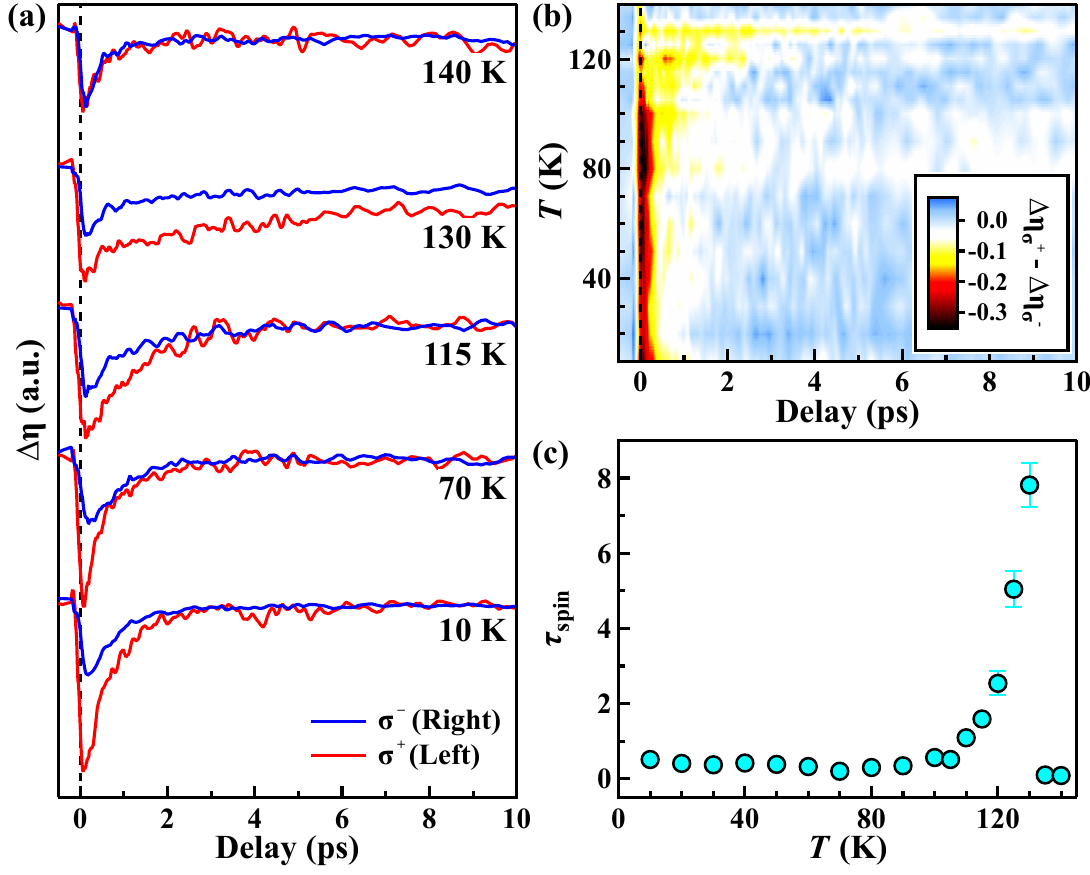}
	\end{center}
	\vspace*{-0.7cm}
	\caption{(a) Transient ellipticity ($\Delta\eta$) as a function of delay time, induced by left ($\sigma^+$) and right ($\sigma^-$) circularly polarized light at selected temperatures. (b) 2D false-color imag of [$\Delta\eta_{\sigma^+}$ - $\Delta\eta_{\sigma^-}$] as a function of temperature and delay time. (c) Temperature dependence of $\tau_{spin}$ extracted using an exponential decay model. Error bars are the standard error in the exponential fitting.}
	\label{FIG:2}
\end{figure}

Beyond the abrupt change observed in the slow component below $T_{\rm SDW}$, a distinct anomaly was noted in the fast component below 115 K. Figures \blue{1(e)} and \blue{1(f)} illustrate the temperature dependence of the amplitude ($A_f$) and relaxation time ($\tau_f$) of this fast component. Both $A_f$ and $\tau_f$ show a steady increase around $T_{\rm SDW}$. However, a significant downturn in $\tau_f$ is observed as the temperature drops below 115 K. This particular trend suggests the onset of an additional DW order, a phenomenon previously documented in transport experiments \cite{RKhasanov2025, YNZhang2024, GWang2024, GWu2001}. While the anomaly in $\tau_f$ serves as a clear indicator for this transition, unlike the divergence of $\tau_s$, it does not allow for a quantitative extraction of the associated energy gap using the RT model. Consequently, the onset temperature of this trend is designated as $T_{\rm DW}$.

\begin{figure*}[t]
	\vspace*{-0.2cm}
	\begin{center}
		\includegraphics[width=2\columnwidth]{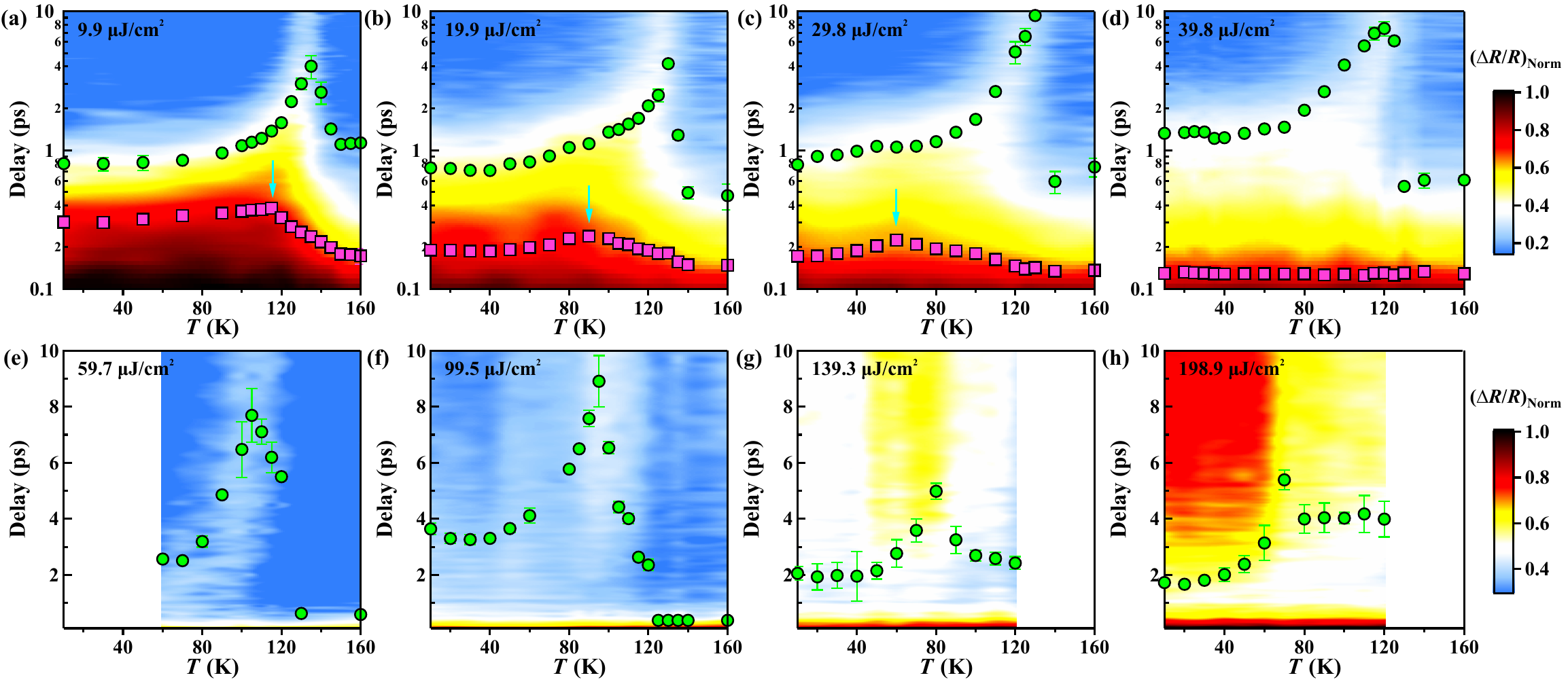}
	\end{center}
	\vspace*{-0.7cm}
	\caption{2D false color image of $\Delta R/R$ as a function of temperature and delay time at various pump fluences: (a) 9.9 {\mjcm}, (b) 19.9 {\mjcm}, (c) 29.8 {\mjcm}, (d) 39.8 {\mjcm}, (e) 59.7 {\mjcm}, (f) 99.5 {\mjcm}, (g) 139.3 {\mjcm}, and (h) 198.9 {\mjcm}. The green scatter points in the top panel represent the extracted $\tau_f$, and the purple scatter points in all figures represent the extracted $\tau_s$. The phonon bottleneck effect is observed at all fluences. Error bars are the standard error in the exponential fitting.}
	\label{FIG:4}
\end{figure*}

To further elucidate the nature of the DW orders in {\LNO}, we investigated the spin relaxation dynamics by conducting transient ellipticity measurements using both left-circularly polarized (LCP, $\sigma^+$) and right-circularly polarized (RCP, $\sigma^-$) light. Figure \blue{2(a)} compares the transient ellipticities excited by LCP ($\Delta\eta_{\sigma^+}$) and RCP ($\Delta\eta_{\sigma^-}$) light across various temperatures. Notably, $\Delta\eta_{\sigma^+}$ and $\Delta\eta_{\sigma^-}$ are nearly identical at 140 K ($T_{\rm SDW}$). However, as the temperature decreases below $T_{\rm SDW}$, a clear difference emerges between $\Delta\eta_{\sigma^+}$ and $\Delta\eta_{\sigma^-}$, with $\Delta\eta_{\sigma^+}$ consistently larger than $\Delta\eta_{\sigma^-}$. Figure \blue{2(b)} presents a 2D false color image illustrating the temperature dependence of the difference signal, $\Delta\eta_{\sigma^+}$ - $\Delta\eta_{\sigma^-}$. An apparent anomaly is discernible near 130 K (slightly below $T_{\rm SDW}$) in this signal. To quantitatively analyze the spin dynamics of {\LNO}, we fitted the difference ($\Delta\eta_{\sigma^+}$ - $\Delta\eta_{\sigma^-}$) using an two-exponential decays (see Sec/ \blue{IV} of the Supplemental Material \cite{SUPPM} for details). The extracted spin relaxation time $\tau_{spin}$ is displayed in Fig. \blue{2(c)}. The temperature dependence of $\tau_{spin}$ closely resembles that of $\tau_s$ [Fig. \blue{1(d)}], both exhibiting divergence around $T_{\rm SDW}$. This strong correlation confirms that the DW order observed in $\Delta R/R$, which is responsible for the gap opening below $T_{\rm SDW}$, is indeed a SDW order. The pronounced difference between the responses to LCP and RCP light below $T_{\rm SDW}$ closely resembles phenomena observed in iron-based superconductors, where nematic fluctuations are known to play a crucial role \cite{APatz2014}. Such a measurement is highly sensitive to the in-plane anisotropy characteristic of nematicity; the breaking of four-fold rotational symmetry induces a differential optical response to the orthogonal linear components of the circularly polarized probe light. Consequently, the difference signal, $\Delta\eta_{\sigma^+}$ - $\Delta\eta_{\sigma^-}$, serves as a direct probe of this broken symmetry. Consistent with our observation, recent resonant soft x-ray scattering and polarimetry studies on thin films of bilayer {\LNO} have revealed anisotropic spin stripe domains indicative of rotational symmetry breaking \cite{NKGupta2025}. Taken together, these findings strongly suggest the presence of electronic nematicity in {\LNO}---analogous to that in cuprate and iron-based superconductors---and point to a close correlation between the SDW and the emergence of superconductivity under high pressure. Conversely, no anomaly in the spin relaxation time is observed near 115 K, indicating that the lower-temperature DW order is likely a spin-independent charge order.

Following the discussion on the nature of DW orders in {\LNO}, the fluence dependence of the transient reflectivity was utilized to further explore the behavior of the DW orders. Figure \blue{3} presents 2D false color image of the normalized $\Delta R/R$ [($\Delta R/R$)$_{\rm Norm}$] as functions of temperature and delay time at various pump fluences. In the top panels [Figs. \blue{3(a)}-\blue{3(d)}], the delay time axis is plotted logarithmically to accentuate the rapid initial dynamics observed at lower fluences. These figures also display the extracted fast ($\tau_f$) and slow ($\tau_s$) relaxation times, denoted by magenta and green scatter points, respectively. As the pump fluence increases, the onset temperature ($T_{\rm DW}$) where $\tau_f$ exhibits an anomaly (marked by cyan arrows) undergoes significant suppression. This suppression causes $T_{\rm DW}$ to reach zero at 39.8 {\mjcm}, beyond which $\tau_f$ remains constant across all temperatures, as evident in Fig. \blue{3(d)}. Conversely, $T_{\rm SDW}$ shows only a slight reduction at pump fluences below roughly 39.8 {\mjcm}. With further increases in pump fluence, $T_{\rm SDW}$ gradually decreases to approximately 70 K, even at the maximum experimental fluence of about 198.9 {\mjcm}, as illustrated in Figs. \blue{3(e)}-\blue{3(h)}. Furthermore, the sign of $A_s$ changes at the temperature where $\tau_s$ reaches its maximum, becoming positive at all temperatures around $\sim$ 139.3 {\mjcm}, as detailed in Sec. \blue{V} of the Supplemental Material \cite{SUPPM}. These characteristics may arise from the partial suppression of the SDW gap induced by the pump fluence. It is important to note that the heating effect induced by the pump fluence is negligible in these experiments. Even under the worst-case scenario---an excitation density of 198.9 {\mjcm}---a steady-state heat diffusion model \cite{JDemsar2006} estimates an average laser-induced temperature rise of only approximately 3 K (see Sec. \blue{VI} of Supplemental Material \cite{SUPPM}).

Figure \blue{4} presents a summary of the extracted SDW gap $\Delta_{\rm SDW}$, $T_{\rm SDW}$, and $T_{\rm DW}$ as a function of the pump fluence, compiled into a comprehensive temperature-fluence phase diagram. Both $T_{\rm SDW}$ and $T_{\rm DW}$ are observed to be suppressed by increasing fluence. The experimentally observed suppression of $T_{\rm DW}$ with increasing fluence is consistent with earlier transport measurements showing that $T_{\rm DW}$ is suppressed under increased pressure \cite{GWang2024, GWu2001, RKhasanov2025}, indicating a similar inhibitory effect of enhanced external fields on $T_{\rm DW}$. In contrast, $T_{\rm SDW}$ decreases slowly as the fluence increases, a trend consistent with observations from recent pump-probe experiments conducted under pressure \cite{YMeng2024}. However, the behavior of $T_{\rm SDW}$ as a function of fluence deviates from the pressure-induced enhancement of $T_{\rm SDW}$ observed in $\mu$SR measurements \cite{KChen2024, RKhasanov2025}, indicating that enhanced external fields have distinct effects on $T_{\rm SDW}$. As $T_{\rm SDW}$ decreases with increasing fluence, the extracted SDW gap size diminishes from approximately 69 meV at 9.9 {\mjcm} to about 15 meV at 198.9 {\mjcm}. We calculated the corresponding ratio of 2$\Delta_{\rm SDW}/k_BT_{\rm SDW}$ at different fluences, as shown in the top panel of Fig. \blue{4}. The ratio of approximately 11 under lower fluence is consistent with the parent compounds of iron-based superconductors \cite{LStojchevska2012, ACharnukha2013, WZHu2008, APogrebna2014}, suggesting a potential similarity between the {\LNO} and iron-based superconductors. Additionally, this ratio gradually decreases with increasing fluence, in line with a decrease of the ratio with increasing doping in iron-based superconductors \cite{LStojchevska2012}. These characteristics collectively indicate that both ultrashort optical pulses, serve as tools for manipulating the intrinsic states of materials \cite{HLiu2025, SDuan2023}, and external pressure are effective in modulating the DW orders of {\LNO}. Unlike previous pump-probe experiments that observed the re-emergence of a DW-II order upon complete suppression of the SDW by high pressure \cite{YMeng2024}, this study did not find such re-emergence, possibly because the SDW was not fully suppressed here.

Our ultrafast optical spectroscopy provides clear evidence that the DW order observed near 140 K in {\LNO} is a SDW order. Furthermore, we note that the relaxation time of the SDW component near $T_{\rm SDW}$ observed in {\LNO} is generally significantly longer than the relaxation time of approximately 1 ps reported in iron-based superconductors \cite{TMertelj2010, LStojchevska2012}. Two potential mechanisms may account for this difference: (i) {\LNO} hosts a greater number of high-energy bosons participating in the bottleneck effect compared to iron-based superconductors, leading to a higher population of quasiparticles being re-excited and consequently prolonging the quasiparticle relaxation time. (ii) In {\LNO}, the processes by which high-energy bosons participating in the bottleneck effect transfer energy into low-energy bosons are slower than in iron-based superconductors, resulting in a longer time required for these high-energy bosons to escape the bottleneck state, thereby significantly extending the duration for which the system maintains a quasi-equilibrium state. In iron-based superconductors, antiferromagnetic magnons are regarded as the high-energy bosons involved in the bottleneck effect, and their smaller number contributes to the significantly reduced relaxation time \cite{XChen2024}. However, recent RIXS experiments in {\LNO} reported spin excitations of approximately 70 meV \cite{XChen2024}, which is much smaller than the SDW gap scale (2$\Delta_{\rm SDW}$). Combined with theoretical calculations indicating that the maximum energy scale of phonons in {\LNO} reaches approximately 100 meV \cite{JZhan2025, ZOuyang2024}, approaching the SDW gap scale, we consider that the high-energy bosons participating in the bottleneck effect in {\LNO} might be phonons rather than magnons. Moreover, we obtained the parameter $g$, which reflects the effective number of the relevant bosons, to be approximately 55 (see Sec. \blue{III} of the Supplemental Material \cite{SUPPM}), according to the RT model. This value is significantly larger than that in iron-based superconductors and comparable to that in cuprate superconductors (see Table \blue{S1} in Sec. \blue{III} of the Supplemental Material \cite{SUPPM}), indicating that the number of phonon degrees of freedom far exceeds the number of magnon degrees of freedom. Therefore, the first mechanism can effectively explain the result that the relaxation time of the SDW component in {\LNO} is significantly longer than that in iron-based superconductors.

\begin{figure}[t]
\vspace*{-0.2cm}
\begin{center}
\includegraphics[width=0.85\columnwidth]{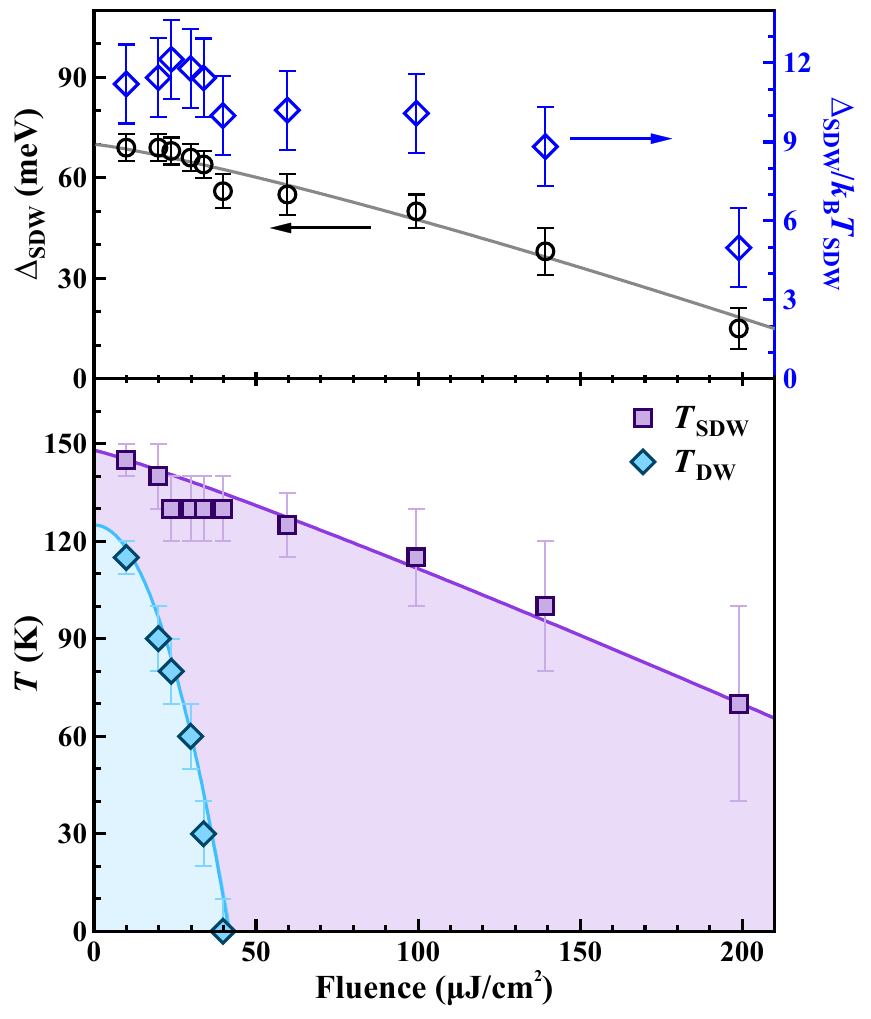}
\end{center}
\vspace*{-0.7cm}
\caption{Temperature-fluence phase diagram of {\LNO}. The upper panel summarizes the SDW gap magnitudes and the ratio of 2$\Delta_{\rm SDW}/T_{\rm SDW}$ as a function of pump fluence, extracted using the RT model. The lower panel presents the onset temperatures of the two DW orders identified in this study. Solid lines serve as guides to the eye.}
\label{FIG:5}
\end{figure}

In cuprates, it is widely known that charge and spin orders are strongly intertwined and compete with superconductivity. Consequently, investigating the existence of charge and spin orders and their relationship with superconductivity in nickelates becomes a key research focus. Indeed, previous studies reported the coexistence of charge and spin orders in the single-layer La$_2$NiO$_4$ and trilayer La$_4$Ni$_3$O$_{10}$ nickelate superconductors \cite{ARicci2021, JZhang2020}. Additionally, recent work observed the coexistence of two distinct orders in {\LNO} \cite{RKhasanov2025}. Our work reconfirms the coexistence of two distinct DW orders in {\LNO}, and further reveals that the lower temperature DW order with $T_{\rm DW}$ $\sim$ 115 K under low pump fluence is irrelevant to the spin degree of freedom. Theoretical calculations predicted the possible existence of short-range CDW orders in {\LNO}, which can be suppressed by the external parameters such as pressure and doping \cite{XJChen2025}. This is consistent with our experimental observation that the lower temperature DW order is significantly suppressed by the increasing pump fluence and eventually vanishes around $\sim$39.8 {\mjcm}. Thus, this nonmagnetic DW order is likely a charge-related DW order.

Finally, we estimated the $e$-$p$h coupling constant ($\lambda$) based on the simple relation $\tau_{e-ph}$ = $\pi$$k_BT_e$$/3\hbar$$\lambda$$\langle\omega^2\rangle$. Here, $\tau_{e-ph}$ represents the $e$-$p$h lifetime, $k_B$ is the Boltzmann constant, $T_e$ is the electron temperature, and $\omega$ is the phonon frequency. Recognizing that DW orders significantly influence the $e$-$p$h lifetime, we selected $\tau_{e-ph}$ values obtained under high fluence and at temperatures $T > T_{\rm SDW}$ to estimate the $e$-$p$h coupling constant. This estimation yielded a value of $\lambda \approx$ 0.21 (detailed calculations can be found in Sec. \blue{VII} of Supplemental Material \cite{SUPPM}). Although our calculated $\lambda$ value significantly exceeds that reported in previous ultrafast spectroscopy studies \cite{YLi2025}, it aligns with theoretical predictions \cite{JZhan2025} but remains insufficient to account for the observed high $T_c$.

In summary, our study provides a comprehensive picture of the competing density waves in {\LNO}. We offer the first direct, dynamic characterization of the 140 K SDW and its associated nematicity using time-resolved ellipticity, and we clearly identify a second, nonmagnetic charge order at 115 K. The most significant advance is our demonstration of selective optical control over these coexisting orders. Our temperature-fluence phase diagram reveals a striking stability hierarchy: the charge order is fragile and easily suppressed by photoexcitation, while the SDW is remarkably robust . This establishes ultrafast optical pulses as a tool to disentangle and manipulate the complex electronic landscape in nickelates, providing a crucial foundation for understanding the interplay between competing orders and high-temperature superconductivity.

This work was supported by the National Natural Science Foundation of China (Grants No. 92265101, No. 12074436, No. 12425404), the National Key Research and Development Program of China (Grant No. 2022YFA1604200), the Open Project of Beijing National Laboratory for Condensed Matter Physics (Grant No. 2024BNLCMPKF001), the Science and Technology Innovation Program of Hunan province (2022RC3068), the Guangdong Basic and Applied Basic Research Funds (Grant No. 2024B1515020040), Guangdong Provincial Key Laboratory of Magnetoelectric Physics and Devices (Grant No. 2022B1212010008), and Research Center for Magnetoelectric Physics of Guangdong Province (2024B0303390001).

\end{document}